\documentclass[preprint,showpacs,preprintnumbers,amsmath,amssymb]{revtex4}

\usepackage{graphicx}
\usepackage{dcolumn}
\usepackage{bm}

\newcommand\GeV{\,\mbox{GeV}}

\begin{document}

\title{NNLO QCD contributions to the flavor nonsinglet sector of ${\bf {F_2(x,Q^2)}}$}
\author{Ali N. Khorramian}%
\email{Khorramiana@theory.ipm.ac.ir}
\homepage{http://particles.ipm.ir/}
\author{S. Atashbar Tehrani}
 \email{Atashbar@ipm.ir}
\affiliation{%
Physics Department, Semnan University, Semnan, Iran}
\affiliation{%
School of Particles and Accelerators,IPM(Institute for Studies in
Theoretical Physics and Mathematics), P.O.Box 19395-5531, Tehran,
Iran}


\date{\today}

\begin{abstract}
We present the results of our QCD analysis for nonsinglet
unpolarized quark distributions and structure function
$F_2(x,Q^2)$.  New parameterizations are derived for the
nonsinglet quark distributions for the kinematic wide range of $x$
and $Q^2$. The analysis is based on the Jacobi polynomials
expansion of the structure function. The higher twist
contributions of proton and deuteron structure function are
obtained in the large $x$ region. Our calculations for nonsinglet
unpolarized quark distribution functions based on the Jacobi
polynomials method are in good agreement with the other
theoretical models. The values of $\Lambda_{QCD}$ and
$\alpha_s(M_z^2)$ are determined.
\end{abstract}

\maketitle

\section{\label{sec:Introduction}Introduction}
The deep-inelastic lepton-nucleon scattering is the source of
important information about the nucleons structure. New and very
precise data on nucleon structure functions have had a profound
impact on our knowledge of parton distributions, in the small and
large $x$ region. During the last years the accuracy of the
obtained experimental data has extensively grown up enough to
study in detail the status of the comparison of the available data
with the theoretical predictions of quantum chromodynamics (QCD)
in the different regions of  momentum transfer.

The importance of deep-inelastic scattering (DIS) for QCD goes
well beyond the measurement of $\alpha_s$ \cite{Altarelli:2008fd}.
In the past it played a crucial role in establishing the reality
of quarks and gluons as partons and in promoting QCD as the theory
of strong interactions. Nowadays it still generates challenges to
QCD as, for example, in the domain of structure functions at small
$x$ \cite{Altarelli:2008aj,Altarelli:2008xp} or of polarized
structure functions \cite{Atashbar Tehrani:2007be} or of
generalized parton densities \cite{Khorramian:2006wg}
 and so on.

 All calculations of high energy processes with initial hadrons, whether
within the standard model or exploring new physics, require parton
distribution functions (PDF's) as an essential input. The
reliability of these calculations, which underpins both future
theoretical and experimental progress, depends on understanding
the uncertainties of the PDF's. The assessment of PDF's, their
uncertainties and extrapolation to the kinematics relevant for
future colliders such as the LHC is an important challenge to high
energy physics in recent years.

 The PDF's are derived
from global analysis of experimental data from a wide range of
hard processes in the framework of perturbative QCD. In this work
this important problem is studied with the help of the method of
the structure function reconstruction over their Mellin moments,
which is based on the expansion of the structure function in terms
of Jacobi polynomials. This method was developed and applied for
QCD analysis
\cite{parisi,Barker,Krivokhizhin:1987rz,Krivokhizhin:1990ct,
Chyla:1986eb,Barker:1980wu,Kataev:1997nc,Kataev:1998ce,Kataev:1999bp,Kataev:2001kk,Khorramian:2007zz}.
The same method has also been applied in polarized case in Refs.
\cite{Leader:1997kw} and \cite{Atashbar
Tehrani:2007be,Khorramian:2007gu,Khorramian:2007zza,Mirjalili:2007ep,Mirjalili:2006hf}.

In this paper we use the deep-inelastic world data for nonsinglet
QCD analysis to obtain the parton distribution function up to
next-to-next-to-leading order (NNLO) approximations. The results
of the present analysis is based on the Jacobi polynomials
expansion of the nonsinglet structure function.

The plan of the paper is to give an introduction of the Jacobi
polynomials approach in Sec.~II. The method of the QCD analysis of
nonsinglet structure function, based on Jacobi polynomials are
written down in this section. In Section~III we present a brief
review of the theoretical formalism of the QCD analysis. A
description of the procedure of the QCD fit of $F_2$ data are
illustrated in Sec.~IV. Section~V contains final results  of the
QCD analysis. Our conclusions are summarized in Sec.~VI.

\section{Jacobi polynomials approach}

The evolution equations allow one to calculate the
$Q^2$-dependence of the parton distributions  provided  at a
certain reference point $Q_0^2$. These distributions are usually
parameterized on the basis of plausible theoretical assumptions
concerning their behavior near the end points $x=0,1$.

One of the simplest and fastest possibilities in the structure
function reconstruction from the QCD predictions for its Mellin
moments is Jacobi polynomials expansion. The Jacobi polynomials
are especially suitable for this purpose since they allow one to
factor out an essential part of the $x$-dependence of the
structure function into the weight function \cite{parisi}. Thus,
given the Jacobi moments $a_{n}(Q^2)$, a structure function
$f(x,Q^2)$ may be reconstructed in a form of the series
\cite{Barker,Krivokhizhin:1987rz,Krivokhizhin:1990ct,Chyla:1986eb,Barker:1980wu}
\begin{equation}
xf(x,Q^2)=x^{\beta}(1-x)^{\alpha} \sum_{n=0}^{N_{max}} a_{n}(Q^2)
\Theta_n ^{\alpha , \beta}(x), \label{e5}
\end{equation}
where $N_{max}$ is the number of polynomials and $\Theta_n
^{\alpha , \beta}(x)$ are the Jacobi polynomials of order $n$,
\begin{equation}
\Theta_{n} ^{\alpha , \beta}(x)= \sum_{j=0}^{n}c_{j}^{(n)}{(\alpha
,\beta )}x^j , \label{e9}
\end{equation}
where $c_{j}^{(n)}{(\alpha ,\beta )}$ are the coefficients
expressed through $\Gamma-$ functions and satisfy the
orthogonality relation with the weight $x^{\beta}(1-x)^{\alpha}$
as in the following:

\begin{equation}
\int_{0}^{1}dx\;x^{\beta}(1-x)^{\alpha}\Theta_{k} ^{\alpha ,
\beta}(x) \Theta_{l} ^{\alpha , \beta}(x)=\delta_{k,l}\
.\label{e8}
\end{equation}
For the moment, we note that the $Q^2$ dependence is entirely
contained in the Jacobi moments
\begin{eqnarray}
a_{n}(Q^2)&=&\int_{0}^{1}dx\;xf(x,Q^2)\Theta_{k} ^{\alpha ,
\beta}(x)\nonumber \\
&=&\sum_{j=0}^{n}\int_{0}^{1}dx\;x^{j+1}c_{j}^{(n)}{(\alpha ,\beta )}f(x,Q^2)\nonumber \\
&=&\sum_{j=0}^{n}c_{j}^{(n)}{(\alpha ,\beta )} f(j+2,Q^2) \;,
\label{e8nn}
\end{eqnarray}
obtained by inverting Eq.~(\ref{e5}), using Eqs.~(\ref{e9},
\ref{e8}) and also definition of moments,
$f(j,Q^2)=\int_{0}^{1}dx\;x^{j-1}f(x,Q^2)$.
\\
Using  Eqs.~(\ref{e5}-\ref{e8nn}) now, one can relate the
structure function with its Mellin moments \begin{eqnarray}
F_{2}^{N_{max}}(x,Q^2)&=&x^{\beta}(1-x)^{\alpha}
\sum_{n=0}^{N_{max}}\Theta_n ^{\alpha, \beta}(x)\nonumber\\
&&\times\sum_{j=0}^{n}c_{j}^{(n)}{(\alpha ,\beta )}
F_{2}(j+2,Q^2), \label{eg1Jacob} \end{eqnarray} where
$F_{2}(j+2,Q^2)$ are the moments determined in the next section.
$N_{max}$, $\alpha$ and $\beta$  have to be chosen so as to
achieve the fastest convergence of the series on the right-hand
side of Eq.~(\ref{eg1Jacob}) and to reconstruct $xg_1$ with the
required accuracy. In our analysis we use $N_{max}=9$,
$\alpha=3.0$ and $\beta=0.5$. The same method has been applied to
calculate the nonsinglet structure function $xF_3$ from their
moments \cite{Kataev:1997nc,Kataev:1998ce,
Kataev:1999bp,Kataev:2001kk} and for polarized structure function
$xg_1$ \cite{Atashbar
Tehrani:2007be,Leader:1997kw,Khorramian:2007gu}.

 Obviously the $Q^2$-dependence
of the polarized structure function is defined by the
$Q^2$-dependence of the moments.

\section{\label{sec:QCD Fit}Theoretical formalism of the QCD analysis}
In the common $\overline{\rm MS}$ factorization scheme the
relevant $F_2$ structure function as extracted from the DIS $ep$
process can be, up to NNLO, written as
\cite{ref3,ref4,ref5,Gluck:2006pm}
\begin{equation}
\label{Eq:F2-tot} F_2(x,Q^2) = F_{2,{\rm NS}}(x,Q^2)+
F_{2,S}(x,Q^2)+ F_{2,g}(x,Q^2)\;.
\end{equation}
The nonsinglet structure function $F_{2,NS}(x,Q^2)$ for three
active (light) flavors has the representation
\begin{eqnarray}
\label{Eq:F2-NS} \frac{1}{x}\, F_{2,{\rm NS}}(x,Q^2) &=&
C_{2,NS}(x,Q^2)\otimes \left[ \frac{1}{18}\, q_8^+
+\frac{1}{6}\, q_3^+\right](x,Q^2)\nonumber \\
&=& \Big[C_{2,q}^{(0)}+aC_{2,{\rm NS}}^{(1)}+a^2C_{2,{\rm
NS}}^{(2)+} \Big]\otimes \left[ \frac{1}{18}\, q_8^+
+\frac{1}{6}\, q_3^+\right](x,Q^2)~.
\end{eqnarray}
The flavor singlet and gluon contributions in
Eq.~(\ref{Eq:F2-tot}) reads
\begin{eqnarray}
\label{Eq:F2-S}\frac{1}{x}\, F_{2,S}(x,Q^2)&=&\frac{2}{9}~C_{2,q}
\otimes \Sigma (x,Q^2)\nonumber
\\&&=\frac{2}{9}  \left[ C_{2,q}^{(0)} +
aC_{2,q}^{(1)}+a^2C_{2,q}^{(2)}\right] \otimes \Sigma (x,Q^2)~;
\end{eqnarray}
\begin{eqnarray}
\label{Eq:F2-g} \frac{1}{x}\,
F_{2,g}(x,Q^2)&=&\frac{2}{9}\;C_{2,g} \otimes g (x,Q^2)
\nonumber \\
&& =\frac{2}{9} \left[aC_{2,g}^{(1)}+a^2C_{2,g}^{(2)}\right]
\otimes g (x,Q^2)~.
\end{eqnarray}

The symbol $\otimes$ denotes the Mellin convolution
\begin{equation}
[A\otimes B](x)=\int_0^1 dx_1\int_0^1
dx_2\;\delta(x-x_1x_2)~A(x_1)B(x_2)\;.
\end{equation}
In Eq.~(\ref{Eq:F2-NS}) $q_3^+=u+\bar{u}-(d+\bar{d})=u_v-d_v$ and
$q_8^+= u+\bar{u}+d+\bar{d}-2(s+\bar{s}) =
u_v+d_v+2\bar{u}+2\bar{d}-4\bar{s}$, where $s=\bar{s}$. Also in
Eq.~(\ref{Eq:F2-S})
 $\Sigma(x,Q^2)\equiv\Sigma_{q=u,d,s}(q+\bar{q})=u_v+d_v+2\bar{u}+2\bar{d}+2\bar{s}$.
Notice that in the above equations
$a=a(Q^2)\equiv\alpha_s(Q^2)/4\pi$ denotes the strong coupling
constant and $C_{i,j}(N)$ are the Wilson coefficients
\cite{Vermaseren:2005qc}.

The combinations of parton densities in the nonsinglet regime and
the valence region $x\geq0.3$ for $F_2^p$ in LO is
\begin{equation}
\frac{1}{x}\,F_2^{p}(x,Q^2) = \left[ \frac{1}{18}\, q_{{\rm
NS,}8}^+ +\frac{1}{6}\, q_{{\rm NS,}3}^+
\right](x,Q^2)+\frac{2}{9} \Sigma(x,Q^2)~,
\end{equation}
where $q_{{\rm NS,}3}^+=u_v-d_v$, $q_{{\rm NS,}8}^+=u_v+d_v$ and
$\Sigma=u_v+d_v$, since sea quarks can be neglected in the region
$x\geq0.3$. So in the $x$-space we have
\begin{eqnarray}
  F_2^{p}(x,Q^2) = \left (\frac{5}{18}\, x\, q_{{\rm NS,}8}^+
  + \frac{1}{6}\, x\, q_{{\rm NS,}3}^+\right) (x,Q^2)& = &
       \frac{4}{9}\,  x\, u_v(x,Q^2)+\frac{1}{9}\, x\, d_v(x,Q^2)~.
\end{eqnarray}
In the above region the combinations of parton densities for
$F_2^d$ are also given by
\begin{eqnarray}
  F_2^{d}(x,Q^2) = \left (\frac{5}{18}\, x\, q_{{\rm NS,}8}^+\right) (x,Q^2)& = &
      \frac{5}{18}\, x(u_v+d_v)(x,Q^2)~,
\end{eqnarray}
where $d=(p+n)/2$ and $q_{{\rm NS,}3}^+=u_v-d_v$.

 In the region $x
\leq 0.3$ for the difference of the proton and deuteron data we
use
\begin{eqnarray}
F_2^{NS}(x,Q^2)&\equiv& 2 (F_2^{p}-F_2^{d})(x,Q^2)\nonumber \\
&=&\frac{1}{3}\, x\, q_{{\rm NS,}3}^+(x,Q^2)= \frac{1}{3}\,
x(u_v-d_v)(x,Q^2)
  +\frac{2}{3}\, x(\bar{u}-\bar{d})(x,Q^2)~,
\end{eqnarray}
where now $q_{{\rm NS,}3}^+ = u_v-d_v+2(\bar{u}-\bar{d})$ since
sea quarks cannot be neglected for $x$ smaller than about 0.3. In
our calculation we supposed the $\bar{d}-\bar{u}$ distribution
\begin{equation}
x(\bar{d}-\bar{u})(x,Q_0^2) = 1.195 x^{1.24}(1-x)^{9.10}
  (1 + 14.05x - 45.52 x^2)~,
  \label{Asy}
\end{equation}
at $Q_0^2=4$ GeV$^2$ which gives a good description of the
Drell-Yan dimuon production data \cite{E866}. In our analysis we
used the above distribution for considering the symmetry breaking
of sea quarks \cite{ref19,Blumlein:2004pr}.
 By using the solution of the
nonsinglet evolution equation for the parton densities to 3$-$
loop order  \cite{Blumlein:2006be}, the nonsinglet structure
functions are given by
\begin{eqnarray}
F_2^k(N,Q^2)&=&\left(1+a\;C_{2,{\rm NS}}^{(1)}(N)+a^2\;C_{2,{\rm
NS}}^{(2)}(N)\right)\times F_2^k(N,Q_0^2)
\left(\frac{a}{a_0}\right)^{-\hat{P}_0(N)/{\beta_0}} \nonumber
\\
&&\Biggl\{1 - \frac{1}{\beta_0} (a - a_0) \left[\hat{P}_1^+(N)
- \frac{\beta_1}{\beta_0} \hat{P}_0(N) \right] \nonumber\\
& & - \frac{1}{2 \beta_0}\left(a^2 - a_0^2\right)
\left[\hat{P}_2^+(N) - \frac{\beta_1}{\beta_0}
\hat{P}_1^+(N)\right. \nonumber \\ &&\left.+ \left(
\frac{\beta_1^2}{\beta_0^2} -
\frac{\beta_2}{\beta_0} \right) \hat{P}_0(N)   \right] \nonumber\\
& & + \frac{1}{2 \beta_0^2} \left(a - a_0\right)^2
\left(\hat{P}_1^+(N) - \frac{\beta_1}{\beta_0} \hat{P}_0(N)
\right)^2 \Bigg\}~.
\end{eqnarray}
Here $k=p,d$ and $NS$ denotes the  three  above cases, i.e.
proton, deuteron and nonsinglet structure function.
$C_{2,NS}^{(m)}(N)$ are the nonsinglet Wilson coefficients in
${\it{O}}(a_s^m)$ which can be found in
\cite{FP,NS2,Vermaseren:2005qc} and $\hat{P}_m$ denote also the
Mellin transforms of the $(m+1)-$ loop splitting functions.

The strong coupling constant $a_{s}$ plays a more central role in
the present paper to the evolution of parton densities. At
$N^{m}LO$ the scale dependence of $a_{s}$ is given by
\begin{eqnarray}
 \label{as-eqn}
  \frac{d\, a_{s}}{d \ln Q^2} \; = \; \beta_{N^mLO}(a_{s})
  \; = \; - \sum_{k=0}^m \, a_{s}^{k+2} \,\beta_k \;.
\end{eqnarray}
The expansion coefficients $\beta_k$ of the $\beta$-function of
QCD are known up to $k=2$, i.e., N$^2$LO
\cite{Tarasov:1980au,Larin:1993tp}

\begin{eqnarray}
 \label{beta-exp}
  \beta_0 &=& 11-2/3~n_f\;,
  \nonumber \\
  \beta_1 &=& 102-38/3~n_f\;,
  \nonumber \\
  \beta_2 &=& 2857/2-5033/18~n_f+325/54~n_f^2\;,
\end{eqnarray}
here $n_f$ stands for the number of effectively massless quark
flavors. The strong coupling constant up to NNLO is as followings
\cite{Vogt:2004ns}:
\begin{eqnarray}
\label{as-exp1}
 a_s(Q^2)
 &=& \frac{1}{\beta_0{L_{\Lambda}}}  -
    \frac{1}{(\beta_0{L_{\Lambda}})^2} b_1\;ln {L_{\Lambda}}  +
    \frac{1}{(\beta_0{L_{\Lambda}})^3} \left[b_1^2 \left(ln^2 {L_{\Lambda}}-ln {L_{\Lambda}}-1
    \right) + b_2 \right],
\end{eqnarray}
where $L_{\Lambda}\equiv ln (Q^2/\Lambda^2)$, $b_k\equiv \beta_k
/\beta_0$, and $\Lambda$ is the QCD scale parameter.

\section{\label{sec:QCD Fit}The  Procedure of the QCD Fits of  $F_2$ Data}

In the present analysis we choose the following parametrization
for the valence quark densities
\begin{eqnarray}
\label{equ:param} x u_v(x,Q^2_0) = {{\cal N}}_u~x^{a_u}(1-x)^{b_u}
(1 + c_u \sqrt{x} + d_u~x)~,\nonumber \\
 x d_v(x,Q^2_0) = {{\cal N}}_d~x^{a_d}(1-x)^{b_d}
(1 + c_d \sqrt{x} + d_d~x)~,
\end{eqnarray}
in the input scale of $Q^2_0=4$ GeV$^2$ and the normalizations
${{\cal N}}_u$ and ${{\cal N}}_d$ being fixed by $\int_0^1 u_v
dx=2$ and $\int_0^1 d_v dx=1$, respectively.  By QCD fits of the
world data for $F_2^{p,d}$, we can extract valence quark densities
using the Jacobi polynomials method. For the nonsinglet QCD
analysis presented in this paper we use the structure function
data measured in charged lepton proton and  deuteron
deep-inelastic scattering. The experiments contributing to the
statistics are BCDMS~\cite{BCDMS}, SLAC~\cite{SLAC},
NMC~\cite{NMC}, H1~\cite{H1}, and ZEUS~\cite{ZEUS}. In our QCD
analysis we use three data samples~: $F_2^p(x,Q^2)$,
$F_2^d(x,Q^2)$ in the nonsinglet regime and the valence quark
region $x \geq 0.3$ and $F_2^{NS} = 2 (F_2^p - F_2^d)$ in the
region $x < 0.3$.

The valence quark region may be parameterized by the nonsinglet
combinations of  parton distributions, which are expressed through
the parton distributions of valence quarks. Only data with $Q^2 >
4~\GeV^2$ were included in the analysis and a cut in the hadronic
mass of $W^2\equiv (\frac{1}{x}-1)\, Q^2+m_{\rm N}^2
> 12.5~\GeV^2$ was applied in order to widely eliminate higher
twist (HT) effects from the data samples. After these cuts we are
left with 762 data points, 322 for $F_2^p$, 232 for $F_2^d$, and
208 for $F_2^{NS}$. By considering the additional cuts on the
BCDMS ($y> 0.35$) and on the NMC data($Q^2 > 8$ GeV$^2$) the total
number of data points available for the analysis reduce from 762
to 551.

The simplest possible choice for the $\chi^2$ function would be
\begin{equation}
\chi^2 = \sum_{i=1}^{n^{data}}
         \frac {(F_{2,i}^{data} - F_{2,i}^{theor})^2}
               {(\Delta F_{2,i}^{data})^2}~,
\end{equation}
where $\Delta F_{2,i}^{data}$ is the error associated with data
point $i$. Through $F_{2,i}^{theor}$, $\chi^{2}$ is a function of
the theory parameters. Minimization of $\chi^{2}$ would identify
parameter values for which the theory fits the data. However, the
simple form is appropriate only for the ideal case of a uniform
data set with uncorrelated errors. For data used in the global
analysis, most experiments combine various systematic errors into
one effective error for each data point, along with the
statistical error. Then, in addition, the fully correlated
normalization error of the experiment is usually specified
separately. For this reason, it is natural to adopt the following
definition for the effective $\chi^2$ \cite{Stump:2001gu}:
\begin{eqnarray}
\chi _{\mathrm{global}}^{2} &=&
\sum_{n} w_{n} \chi _{n}^{2}\;,\qquad (n\;%
\mbox{labels the different experiments}) \label{eq:Chi2global}
\\
\chi _{n}^{2} &=&\left(\frac{1-{\cal N}_{n}}{\Delta{\cal
N}_{n}}\right)^{2} +\sum_{i}\left( \frac{{\cal
N}_{n}F_{2,i}^{data}-F_{2,i}^{theor}}{{\cal N}_{n}\Delta
F_{2,i}^{data}} \right)^{2}\;. \label{eq:Chi2n}
\end{eqnarray}

For the $n^{\mathrm{th}}$ experiment, $F_{2,i}^{data}$, $\Delta
F_{2,i}^{data}$, and $%
F_{2,i}^{theor}$ denote the data value, measurement uncertainty
(statistical and systematic combined), and theoretical value for
the $i^{\mathrm{th}}$ data point. ${\Delta{\cal N}_{n}}$ is the
experimental normalization uncertainty and ${\cal N}_{n}$ is an
overall normalization factor for the data of experiment $n$. The
factor $w_{n}$ is a possible weighting factor(with default value$-
1$).  However, we allowed for a relative normalization shift
${\cal N}_{n}$ between the different data sets within the
normalization uncertainties ${\Delta{\cal N}_{n}}$ quoted by the
experiments. For example the normalization uncertainty of the
NMC(combined) data is estimated to be 2.5\%.
 The normalization shifts ${\cal N}_{n}$ were fitted
once and then kept fixed.

The number of data points for the nonsinglet QCD analysis with
their $x$ and $Q^2$ ranges, and the normalization shifts
determined are summarized in Table.~I. In this table the first
column  gives (in parentheses) the beam momentum in GeV of the
respective data set (number), a flag whether the data come from a
combined analysis of all beam momenta (comb) or whether the data
are taken at high momentum transfer (hQ2). The $x$ and $Q^2$ range
indicate in the second and third columns, respectively. The fourth
column ($F_2$) contains the number of data points according to the
cuts: $Q^2
> 4~{\GeV^2}$, $W^2 > 12.5~{ \GeV^2}$, $x > 0.3$ for $F_2^p$ and
$F_2^d$ and $x < 0.3$ for $F_2^{NS}$. The reduction of the number
of data points by the additional cuts on the BCDMS data ($y >
0.3$) and on the NMC data ($Q^2 > 8~{ \GeV^2}$) are given in the
fifth column ($F_2~cuts$). The last column (${\cal{N}}$) contains
the normalization shifts.

Now the sums in $\chi _{\mathrm{global}}^{2}$ run over all data
sets and in each data set over all data points. The minimization
of the  above $\chi^2$ value to determine the best parametrization
of the unpolarized parton distributions is done using the program
{\tt MINUIT} \cite{MINUIT}.
\renewcommand{\arraystretch}{0.7}
\begin{table*}
\begin{tabular}{||l|    c   | c |c |c ||c||}
\hline
Experiment  & $x$ & $Q^2,~\GeV^2$ & $F_2^p$ & $F_2^p~cuts$ &  ${\cal N}$  \\
\hline  BCDMS (100)   & 0.35 -- 0.75 &  11.75 --  75.00  & 51 & 29
&
1.005 \\
BCDMS (120)   & 0.35 -- 0.75 &  13.25 --  75.00  &  59 &  32 &
0.998 \\
BCDMS (200)   & 0.35 -- 0.75 &  32.50 -- 137.50  &  50 &  28 &
0.998 \\
BCDMS (280)   & 0.35 -- 0.75 &  43.00 -- 230.00  &  49 &  26 &
0.998 \\
NMC (comb)    & 0.35 -- 0.50 &   7.00 --  65.00  &  15 &  14 &
1.000 \\
SLAC (comb)   & 0.30 -- 0.62 &   7.30 --  21.39  &  57 &  57 &
1.013 \\
H1 (hQ2)      & 0.40 -- 0.65 &    200 --  30000  &  26 &  26 &
1.020 \\
ZEUS (hQ2)    & 0.40 -- 0.65 &    650 --  30000  &  15 &  15 &
1.007 \\
\hline
{\bf proton}      &              &                   & 322 & 227 &  \\
\hline
\end{tabular}
\\
\vspace{0.1cm} \text{(a) Number of $F_2^p$ data points.}\\
\begin{tabular}{||l|c | c |c |c ||c||}
\hline
Experiment  & $x$ & $Q^2, \GeV^2$ & $F_2^d$ & $F_2^d~cuts$ & ${\cal N}$  \\
\hline
BCDMS (120)   & 0.35 -- 0.75 & 13.25 --  99.00  &  59 &  32 &    1.001 \\
BCDMS (200)   & 0.35 -- 0.75 & 32.50 -- 137.50  &  50 &  28 &    0.998 \\
BCDMS (280)   & 0.35 -- 0.75 & 43.00 -- 230.00  &  49 &  26 &    1.003 \\
NMC (comb)    & 0.35 -- 0.50 &  7.00 --  65.00  &  15 &  14 &    1.000 \\
SLAC (comb)   & 0.30 -- 0.62 & 10.00 --  21.40  &  59 &  59 &  0.990 \\
\hline
{\bf deuteron}    &              &                  & 232 & 159 &  \\
\hline
\end{tabular}
\\
\vspace{0.1cm} \text{(b) Number of $F_2^d$ data points.}\\
\begin{tabular}{||l|    c   | c |c |c ||c||}
 \hline  Experiment  & $x$ & $Q^2, \GeV^2$ & $F_2^{NS}$ &
$F_2^{NS}~cuts$
& ${\cal N}$  \\
\hline
BCDMS (120)  & 0.070 -- 0.275 &  8.75 --  43.00  &  36 &  30 &     0.983 \\
BCDMS (200)  & 0.070 -- 0.275 & 17.00 --  75.00  &  29 &  28 &     0.999 \\
BCDMS (280)  & 0.100 -- 0.275 & 32.50 -- 115.50  &  27 &  26 &     0.997 \\
NMC (comb)   & 0.013 -- 0.275 &  4.50 --  65.00  &  88 &  53 &     1.000 \\
SLAC (comb)  & 0.153 -- 0.293 &  4.18 --   5.50  &  28 &  28 &     0.994 \\
\hline
{\bf nonsinglet} &               &                  & 208 & 165 &    \\
\hline
\end{tabular}
\\
\vspace{0.1cm} \text{(c) Number of $F_2^{NS}$ data
points.\vspace{0.5cm}} \caption{\label{tab:table3}{\sf Number of
experimental data points (a) $F_2^{p}$, (b) $F_2^{d}$, and (c)
$F_2^{NS}$ for the nonsinglet QCD analysis with their $x$ and
$Q^2$ ranges. The name of different data set
 and range of $x$ and $Q^2$ are given in the three first columns . The fourth column ($F_2$) contains the number of data
points according to the cuts: $Q^2
> 4~{\GeV^2}$, $W^2 > 12.5~{ \GeV^2}$, $x > 0.3$ for $F_2^p$ and
$F_2^d$ and $x < 0.3$ for $F_2^{NS}$. The reduction of the number
of data points by the additional cuts (see text) are given in the
5th column ($F_2~cuts$). The normalization shifts are listed in
the last column.}}
\end{table*}

The  one $\sigma$ error for the parton density $f_q$ as given by
Gaussian error propagation is \cite{Blumlein:2006be}
\begin{eqnarray}
\sigma( f_q(x))^2 =  \sum_{i=1}^{n_p}\sum_{j=1}^{n_p}
                \left( \frac{\partial f_q}{\partial p_i}\right)
                \left(\frac{\partial f_q}{\partial p_j} \right)
                \textrm{cov}(p_i,p_j)~,
\end{eqnarray}
 where the sum runs over all fitted parameters. The functions
$\partial f_q /
\partial p_i$ are the derivatives of $f_q$ with respect to the fit parameter $p_i$,
and $\textrm{cov}(p_i,p_j)$ are the elements of the covariance
matrix. The derivatives $\partial f_q / \partial p_i$ can be
calculated analytically at the input scale $Q_0^2$. Their values
at $Q^2$ are given by evolution which is performed in {Mellin-$N$}
space.

\section{Results}
\begin{figure}
\vspace{4 cm}
\includegraphics[angle=0, width=10.0cm]{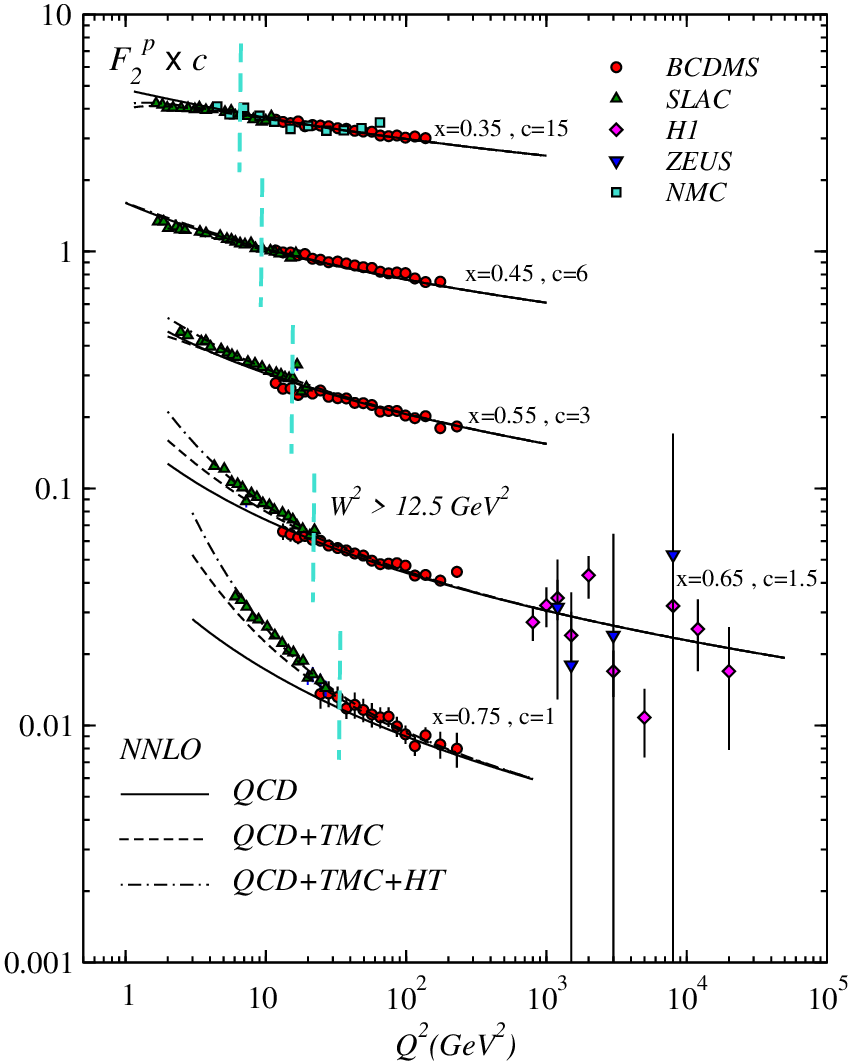}
\caption{\label{fig:1}{\sf The structure function $F_2^p$ as a
function of $Q^2$ in intervals of $x$. Shown are the pure QCD fit
in NNLO (solid line) and the contributions from target mass
corrections (dashed line) and higher twist (dashed--dotted line).
The vertical dashed line indicate the regions with $W^2 >
12.5~{\GeV^2}$.} }
\end{figure}

In the QCD analysis of the present paper we used three data sets:
the structure functions $F_2^{p}(x,Q^2)$  and $F_2^{d}(x,Q^2)$ in
the region  of $x \geq 0.3$ and the combination of these structure
functions $F_2^{\rm NS}(x,Q^2)$ in the region  of $x < 0.3$~.
Notice that we take into account the cuts $Q^2>4$ GeV$^2$,
$W^2>12.5$ GeV$^2$ for our QCD fits to determine some unknown
parameters. In Fig.(\ref{fig:1}) the proton data for $F_2(x,Q^2)$
are shown in the nonsinglet regime and the valence quark region $x
\geq 0.3$ indicating the above cuts by a vertical dashed line. The
solid lines correspond to the NNLO QCD fit.

Now, it is possible to take into account the target mass effects
in our calculations. The perturbative form of the moments is
derived under the assumption that the mass of the target hadron is
zero (in the limit $Q^{2}\rightarrow\infty$). At intermediate and
low $Q^{2}$ this assumption will begin to break down and the
moments will be subject to potentially significant power
corrections, of order ${{\cal O}}~(m_{N}^{2}/Q^{2})$, where $m_N$
is the mass of the nucleon. These are known as target mass
corrections (TMCs) and when included, the moments of flavor
nonsinglet structure function have the form
\cite{Georgi:1976ve,Gluck:2006yz}

\begin{eqnarray}
\label{equ:TMC} F_{2,{\rm TMC}}^{k}(n,Q^2) & \equiv &
  \int_0^1 x^{n-2} F_{2,{\rm TMC}}^{k}(x,Q^2)\, dx  \nonumber\\
& = &F_{2}^{
k}(n,Q^2)+\frac{n(n-1)}{n+2}\left(\frac{m_N^2}{Q^2}\right)\,F_{2}^{
k}(n+2,Q^2)\nonumber \\
&&+\frac{(n+2)(n+1)n(n-1)}{2(n+4)(n+3)}\left(\frac{m_N^2}{Q^2}\right)^2\,F_{2}^{
k}(n+4,Q^2)+ {\cal{O}}\left( \frac{m_N^2}{Q^2}\right)^3~,
\end{eqnarray}
where higher powers than $(m_{\rm N}^2/Q^2)^2$ are negligible for
the relevant $x <0.8$ region. By inserting Eq.~(\ref{equ:TMC}) in
Eq.~(\ref{eg1Jacob}) we have
\begin{eqnarray}
F_{2}^{N_{max},k}(x,Q^2)&=&x^{\beta}(1-x)^{\alpha}
\sum_{n=0}^{N_{max}}\Theta_n ^{\alpha,
\beta}(x)\times\sum_{j=0}^{n}c_{j}^{(n)}{(\alpha ,\beta )}
F_{2,{\rm TMC}}^{k}(j+2,Q^2)\;, \label{eg1JacobTMC} \end{eqnarray}
where $F_{2,{\rm TMC}}^{k}(j+2,Q^2)$ are the moments determined by
Eq.~(\ref{equ:TMC}). In Fig.(\ref{fig:1})  the dashed lines
correspond to the NNLO QCD fit adding target mass corrections.

\begin{figure}
\vspace{4 cm}
\includegraphics[angle=0, width=10.0cm]{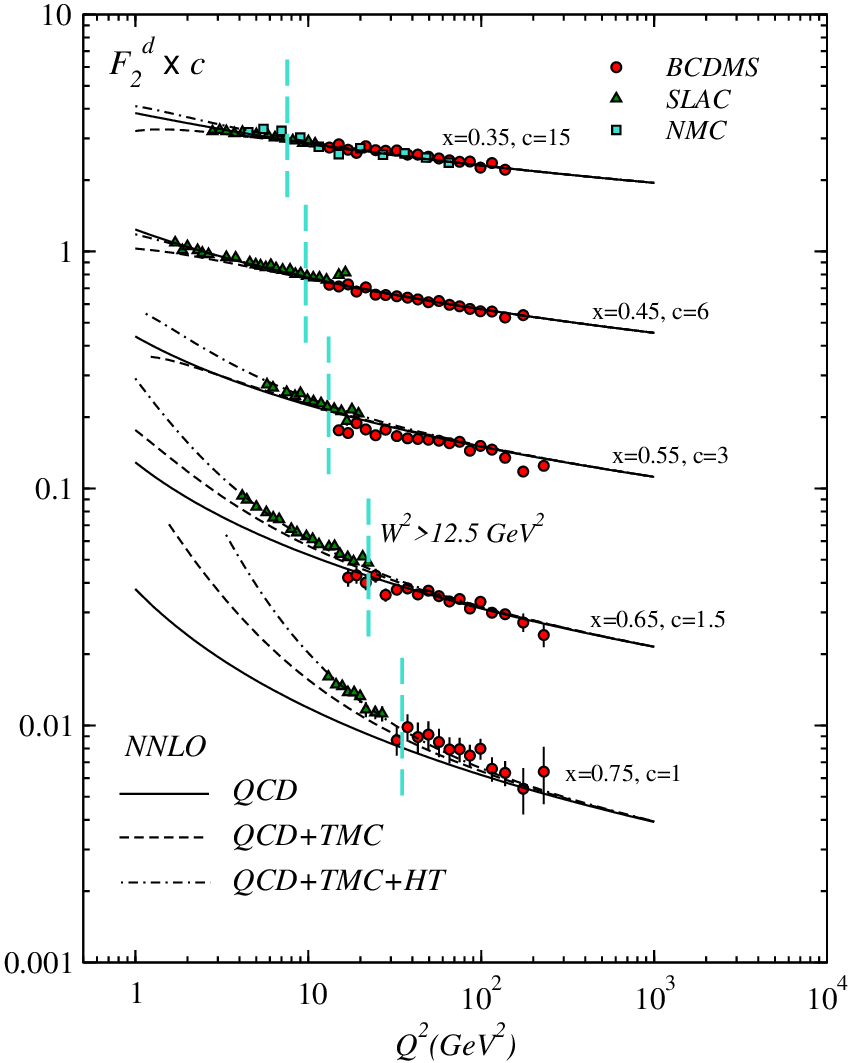}
\caption{\label{fig:2} {\sf The structure function $F_2^d$ as a
function of $Q^2$ in intervals of $x$. Shown are the pure QCD fit
in NNLO (solid line) and the contributions from target mass
corrections  (dashed line) and higher twist (dashed--dotted line).
The vertical dashed lines indicate the regions with $W^2 >
12.5~{\GeV^2}$.} }
\end{figure}

\begin{figure}
\vspace{4 cm}
\includegraphics[angle=0, width=10.0cm]{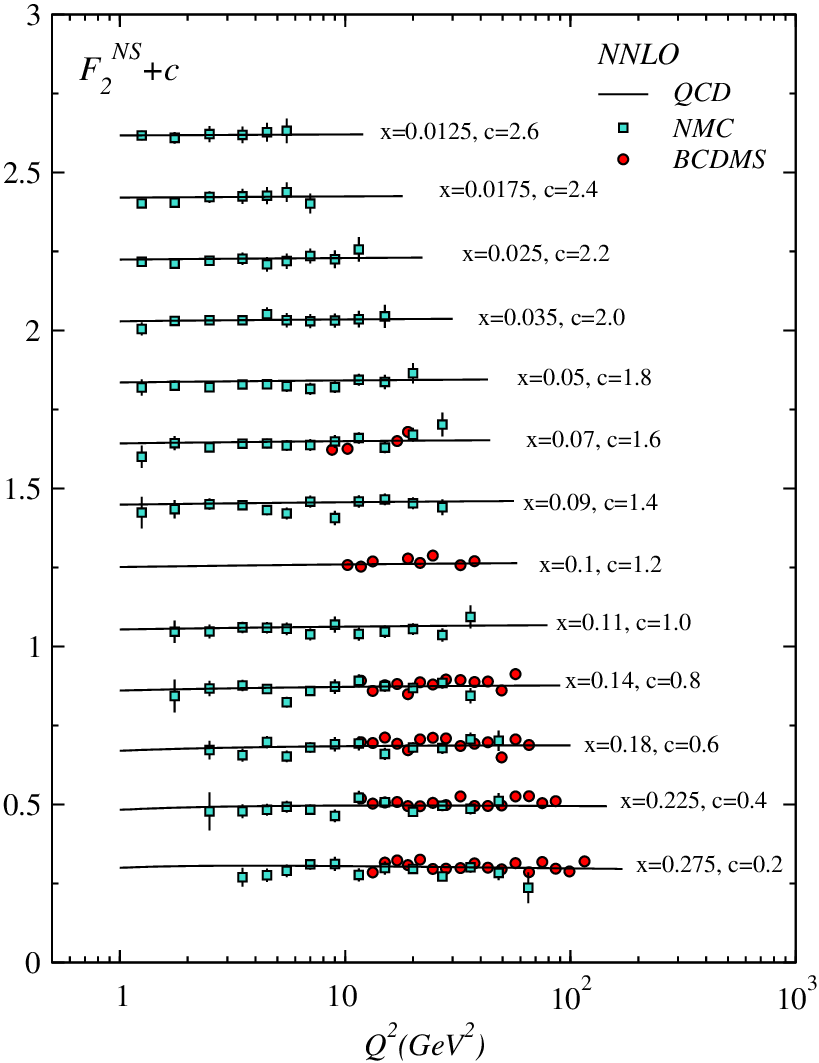}
\caption{\label{fig:3} {\sf The structure function $F_2^{\rm NS}$
as a function of $Q^2$ in intervals of $x$. Shown is the pure QCD
fit in NNLO (solid lines).} }
\end{figure}

Despite the kinematic cuts ($Q^2\geq 4$ GeV$^2$, $\, W^2\equiv
(\frac{1}{x}-1)\, Q^2+m_{\rm N}^2\geq 12.5$ GeV$^2$) used for our
analysis, we also take into account higher twist corrections to
$F_2^{p}(x,Q^2)$ and $F_2^{d}(x,Q^2)$ in the kinematic region $Q^2
\geq 4 \GeV^2, 4<W^2 < 12.5 \GeV^2$ in order to learn whether
nonperturbative effects may still contaminate our perturbative
analysis. For this purpose we extrapolate the QCD fit results
obtained for $W^2 \geq 12.5 \GeV^2$ to the region $Q^2 \geq 4
\GeV^2, 4<W^2 < 12.5 \GeV^2$ and form the difference between data
and theory, applying target mass corrections in addition. Now by
considering higher twist correction (HT)
\begin{eqnarray}
\label{equ:HTC} F_2^{\rm exp}(x,Q^2) = O_{\rm TMC}[F_2^{\rm
HT}(x,Q^2)] \cdot \left( 1 +
\frac{h(x,Q^2)}{Q^2[\GeV^2]}\right)\;,
\end{eqnarray}
the higher twist coefficient can be extract. Here the operation
$O_{\rm TMC}[...]$ denotes taking the target mass corrections of
the twist--2 contributions to the respective structure function.
The coefficients $h(x,Q^2)$ are determined in bins of $x$ and
$Q^2$ and are then averaged over $Q^2$.  We extrapolate our QCD
fits to the region $12.5 \GeV^2 \geq W^2 \geq 4 \GeV^2$ in
Fig.(\ref{fig:1}). The dashed-dotted lines in this figure
  correspond to the NNLO QCD fit adding target mass and higher twist corrections.
 There, at higher values of $x$ a clear gap between the
data and the QCD fit is seen. Figure (\ref{fig:2}) shows the
corresponding results for the deuteron data.  Figure (\ref{fig:3})
shows the result of the pure QCD fit for the nonsinglet structure
function in NNLO.
\begin{table*}
\begin{tabular}{|c|c|c|c|c|}
\hline \hline
           &             & LO  & NLO & NNLO \\
\hline $u_v$      & $a_u$         &  0.6698 $\pm$ 0.0073 &  0.7434
$\pm$ 0.009 &
 0.7772 $\pm$ 0.009 \\
           & $b_u$         &  3.5104 $\pm$ 0.042 &  3.8907 $\pm$ 0.040 &
 4.0034 $\pm$ 0.033 \\
           & $c_u  $    &  0.1990             &  0.1620             &
 0.1000             \\
           & $d_u$    & 1.498             & 1.2100             &
1.1400             \\
\hline $d_v$      & $a_d$         &  0.6850 $\pm$ 0.035 &  0.7369
$\pm$ 0.040 &
 0.7858 $\pm$ 0.043 \\
           & $b_d$         &  3.1685 $\pm$ 0.192 &  3.5051 $\pm$ 0.225 &
 3.6336 $\pm$ 0.244 \\
           & $c_d  $    & 0.5399             & 0.3899             &
0.1838             \\
           & $d_d$    & -1.4000             & -1.3700             &
-1.2152            \\
\hline \multicolumn{2}{|c|}{$\Lambda_{\rm QCD}^{\rm N_f=4}$, MeV}
& 213.2$\pm$ 28 &
263.8 $\pm$ 30 & 239.9 $\pm$ 27 \\
\hline \hline \multicolumn{2}{|c|}{$\chi^2 / ndf$} & 538/546 =
0.9853 & 523/546 = 0.9578 &
506/546 = 0.9267 \\
\hline \hline
\end{tabular}
\caption{\label{tab:table4}{\sf Parameter values of the LO, NLO
and NNLO nonsinglet QCD fit at $Q_0^2 = 4~ \mbox{GeV}^2$. The
values without error have been fixed after a first minimization
since the data do not constrain these parameters well enough (see
text).}}
\end{table*}
\vspace{1 cm}

%

\renewcommand{\arraystretch}{0.9}
\begin{table*}
\begin{tabular}{|c||c|c|c|c|c|}
\hline \hline
{\bf LO} & $a_{u}$ & $b_{u}$ & $a_{d}$ & $b_{d}$& $\Lambda_{\rm QCD}^{\rm N_f=4}$ \\
\hline
 $a_{u}$ &~{\bf 5.28$\times$10$^{-5}$}&  &  &  &  \\
\hline
 $b_{u}$        &~1.65$\times$10$^{-4}$ &{\bf ~1.73$\times$10$^{-3}$} &  &  &  \\
\hline
 $a_{d}$        &~-7.39$\times$10$^{-5}$ &~-4.55$\times$10$^{-4}$ & {\bf ~1.23$\times$10$^{-3}$} &  & \\
\hline
 $b_{d}$        &~-2.64$\times$10$^{-4}$ &~-2.12$\times$10$^{-3}$ &~6.15$\times$10$^{-3}$ &~{\bf 3.67$\times$10$^{-2}$} & \\
\hline
 $\Lambda_{QCD}^{(4)}$        &~1.90$\times$10$^{-5}$ &~-8.34$\times$10$^{-4}$ &~2.39$\times$10$^{-5}$ ~&-3.16$\times$10$^{-4}$ &~{\bf 7.79$\times$10$^{-4}$} \\
\hline \hline
{\bf NLO}  & $a_{u}$ & $b_{u}$ & $a_{d}$ & $b_{d}$& $\Lambda_{\rm QCD}^{\rm N_f=4}$ \\
\hline
 $a_{u}$ & {\bf ~8.87$\times$10$^{-5}$} &  &  &  &  \\
\hline
 $b_{u}$        &~2.39$\times$10$^{-4}$ &~{\bf 1.63$\times$10$^{-3}$} &  &  &  \\
\hline
 $a_{d}$        &~-1.34$\times$10$^{-4}$ &~-7.86$\times$10$^{-4}$ &~{\bf 1.61$\times$10$^{-3}$} &  & \\
\hline
 $b_{d}$        &~-5.10$\times$10$^{-4}$ &~-4.19$\times$10$^{-3}$ &~8.33$\times$10$^{-3}$ &~{\bf 5.07$\times$10$^{-2}$} & \\
\hline
 $\Lambda_{QCD}^{(4)}$        &~8.71$\times$10$^{-5}$ &~-5.39$\times$10$^{-4}$& 8.09$\times$10$^{-5}$ &~2.57$\times$10$^{-4}$ &~{\bf 8.80$\times$10$^{-4}$} \\
\hline \hline
{\bf NNLO}  & $a_{u}$ & $b_{u}$ & $a_{d}$ & $b_{d}$& $\Lambda_{\rm QCD}^{\rm N_f=4}$ \\
\hline
 $a_{u}$ &~{\bf 7.61$\times$10$^{-5}$} &  &  &  &  \\
\hline
 $b_{u}$        &~1.73$\times$10$^{-4}$ &~{\bf 1.10$\times$10$^{-3}$} &  &  &  \\
\hline
 $a_{d}$        &~-8.41$\times$10$^{-5}$ &~-6.62$\times$10$^{-4}$ &~{\bf 1.85$\times$10$^{-3}$} &  & \\
\hline
 $b_{d}$        &~-2.73$\times$10$^{-4}$ &~-3.73$\times$10$^{-3}$ &~9.79$\times$10$^{-3}$ &~{\bf 5.98$\times$10$^{-2}$} & \\
\hline
 $\Lambda_{QCD}^{(4)}$        &~1.08$\times$10$^{-4}$ &~-2.74$\times$10$^{-4}$ &~1.06$\times$10$^{-4}$ &~4.19$\times$10$^{-4}$ &~{\bf 7.41$\times$10$^{-4}$} \\
\hline \hline
\end{tabular}

\caption{\label{tab:table5}{\sf Our results for the covariance
matrix of the  LO, NLO, and NNLO nonsinglet QCD fit at $Q_0^2 =
4~\GeV^2$ by using MINUIT\cite{MINUIT}. }}
\end{table*}

In Table (\ref{tab:table4}) we summarize the LO, NLO, and NNLO fit
results without HT contributions for the parameters of the parton
densities $xu_v(x,Q^2_0)$, $xd_v(x,Q^2_0)$ and $\Lambda_{\rm
QCD}^{\rm N_f =4}$. The resulted value of $\chi^2/ndf$ is 0.9853
at LO, 0.9578 at NLO, and 0.9267 at NNLO. Our results for
covariance matrix for LO, NLO, and NNLO are presented in
Table(\ref{tab:table5}).

\begin{figure}
\vspace{4 cm}
\begin{center}
\includegraphics[angle=0, width=8.0cm]{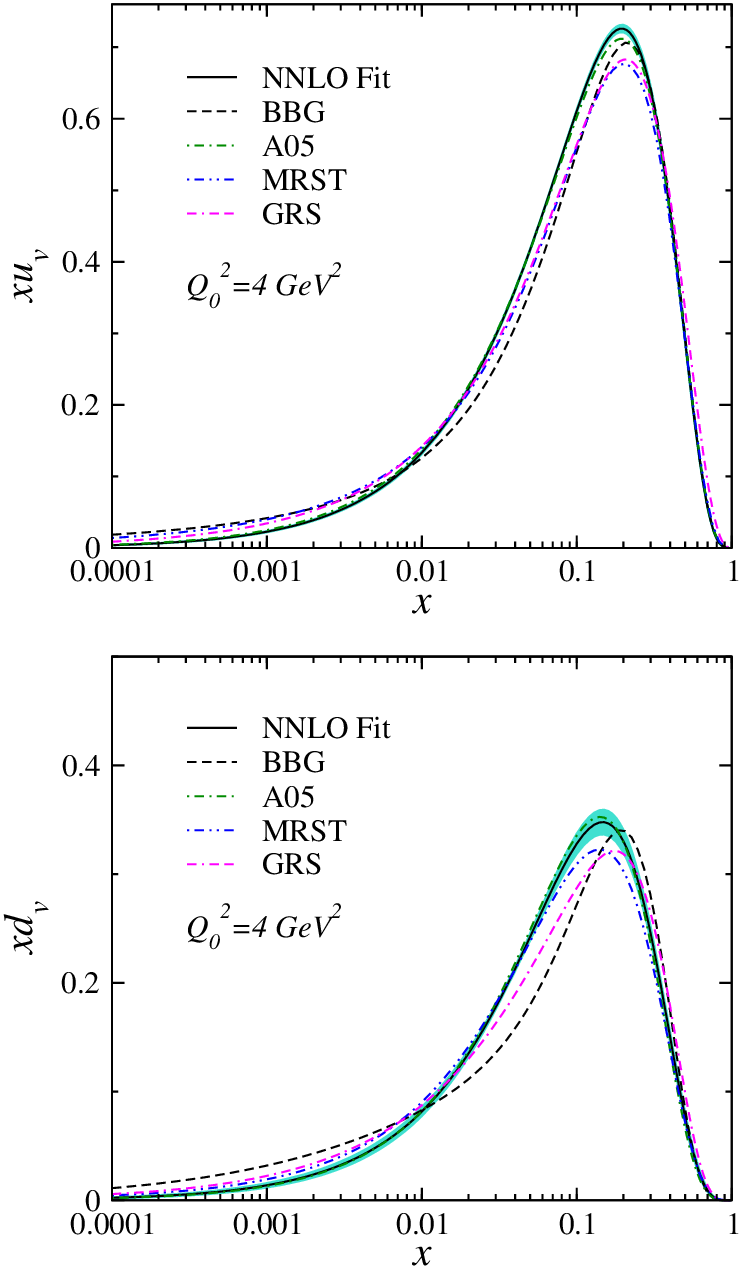}
\end{center}
{\sf \caption{\label{fig:4} {\sf The parton densities $xu_v$ and
$xd_v$ at the input scale $Q_0^2 = 4.0~{\rm{\GeV^2}}$ (solid line)
compared with results obtained from NNLO analysis by BBG (dashed--
line)~\cite{Blumlein:2006be}, A05 (dashed--dotted
line)~\cite{Alekhin:2005gq}, MRST (dashed--dotted--dotted
line)~\cite{Martin:2007bv}, and GRS (dashed--dashed--dotted
line)~\cite{Gluck:2006yz}. The shaded areas represent the fully
correlated one $\sigma$ statistical error bands.}}}
\end{figure}

Figure (\ref{fig:4}) illustrates our fit results for
$xu_v(x,Q^2_0)$, $xd_v(x,Q^2_0)$ at $Q_0^2 = 4 \GeV^2$ at NNLO
with correlated errors. We compare with the results of
\cite{Blumlein:2006be,Alekhin:2005gq,Gluck:2006yz} and a very
recent analysis \cite{Martin:2007bv}. Our results for
$xu_v(x,Q^2_0)$ and $xd_v(x,Q^2_0)$ are in good agreement with the
other theoretical model at the one $\sigma$ level.

\begin{figure}
\vspace{4 cm}
\includegraphics[angle=0, width=12cm]{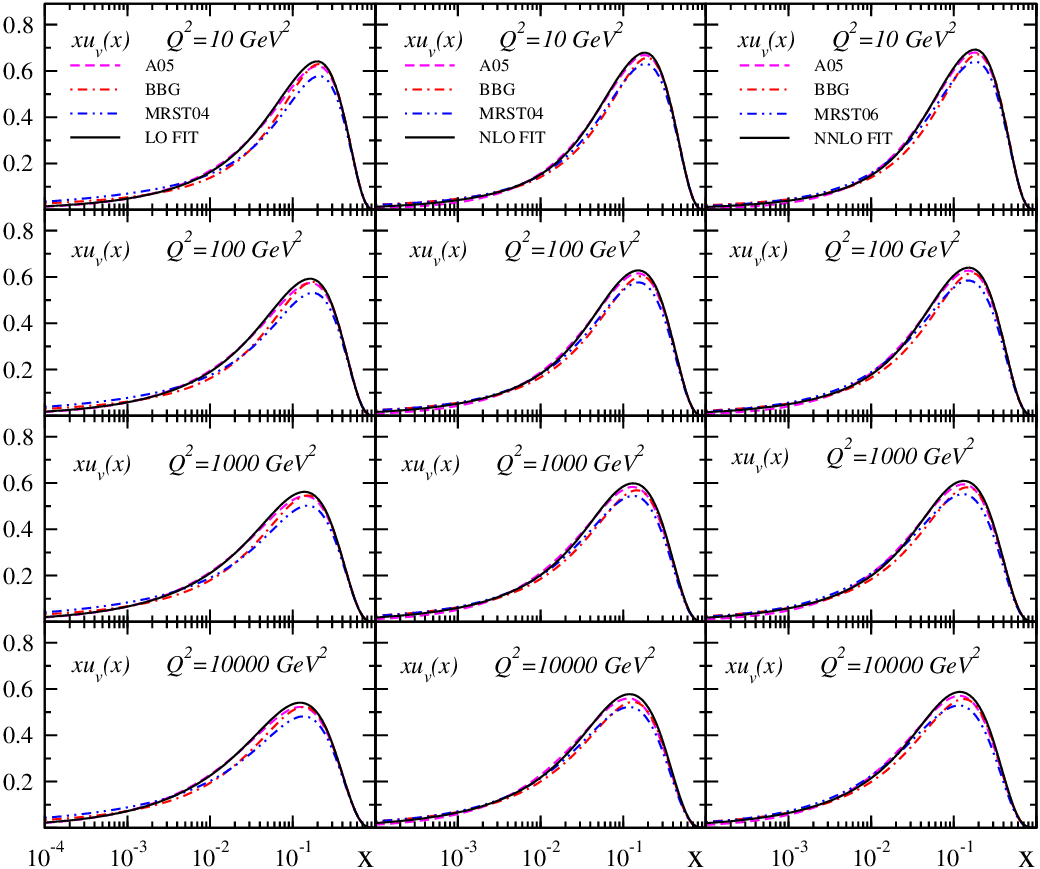}
\begin{center}
\end{center}
{\sf \caption{\label{fig:5} {\sf The parton density $xu_v$ at NNLO
evolved up to $Q^2 = 10,000~{ \GeV^2}$ (solid lines) compared with
results obtained by A05 (dashed line)~\cite{Alekhin:2005gq}, BBG
(dashed--dotted line)~\cite{Blumlein:2006be}, and MRST
(dashed--dotted-dotted line)~\cite{Martin:2007bv,MRST04}.}}}
\end{figure}

\begin{figure}
\vspace{4 cm}
\includegraphics[angle=0, width=12cm]{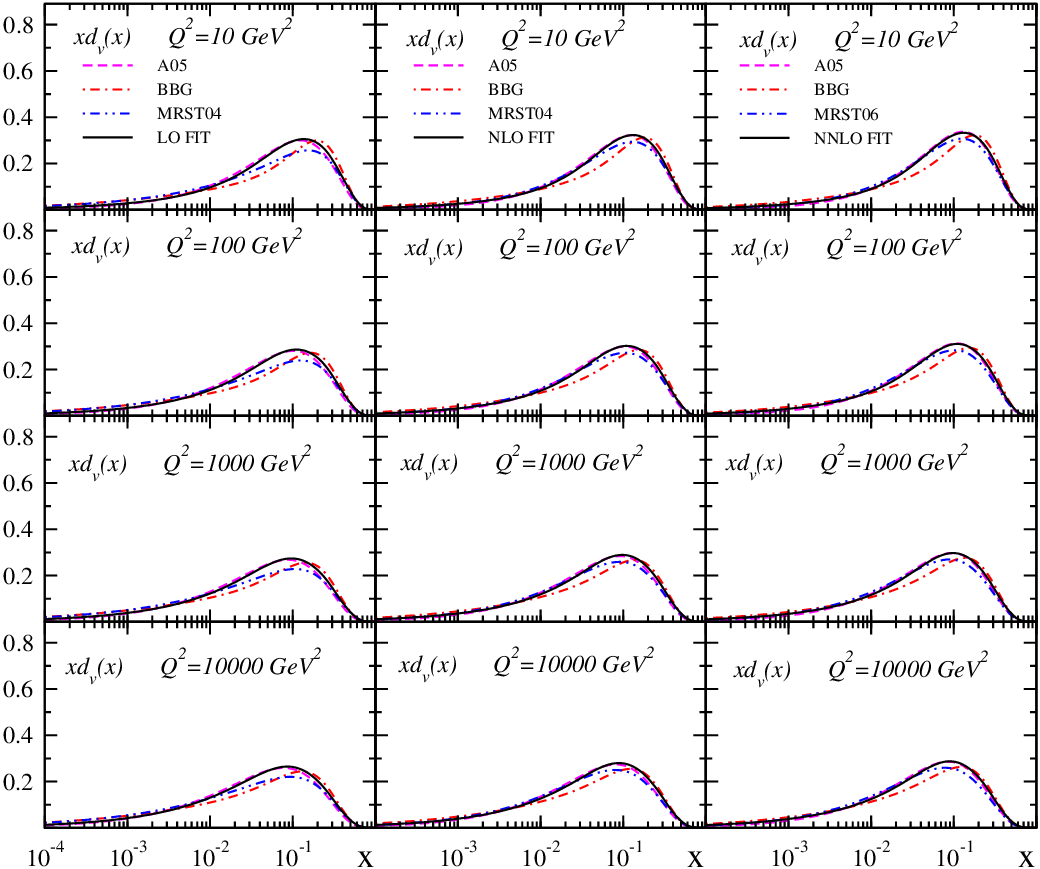}
\begin{center}
\end{center}
{\sf \caption{\label{fig:6} {\sf The parton density $xd_v$ at NNLO
evolved up to $Q^2 = 10,000~{ \GeV^2}$ (solid lines) compared with
results obtained by A05 (dashed line)~\cite{Alekhin:2005gq}, BBG
(dashed--dotted line)~\cite{Blumlein:2006be}, MRST
(dashed--dotted--dotted line)~\cite{Martin:2007bv,MRST04}.}}}
\end{figure}

In Figs.~(\ref{fig:5}) and (\ref{fig:6}) we show the evolution of
the valence quark distributions $xu_v(x,Q^2)$ and $xd_v(x,Q^2)$
from $Q^2 = 10 \GeV^2$ to $Q^2 = 10^4 \GeV^2$ in the region  $x\in
[10^{-4},1]$ up to NNLO. We also compared with other QCD analysis
\cite{Alekhin:2005gq,Blumlein:2006be,Martin:2007bv,MRST04}. With
rising values of $Q^2$ the distributions flatten at large values
of $x$ and rise at low values.

Another way to compare the NNLO fit results consists in forming
moments of the distributions $u_v(x,Q^2), d_v(x,Q^2),$ and
$u_v(x,Q^2)- d_v(x,Q^2)$. In Table~\ref{tab:table6} we present the
lowest non-trivial moments of these distributions at $Q^2 = Q_0^2$
in NNLO and compare to the respective moments obtained for the
parameterizations
\cite{Blumlein:2006be,MRST04,A02,Alekhin:2006zm}.
\renewcommand{\arraystretch}{1.1}
\begin{table*}
\begin{tabular}{|c|c|c|c|c|c|c|}
\hline \hline
 $~f~$ & $~~N~~$ & NNLO & BBG & MRST04 &~~~A02~~~&~~~A06~~~\\
\hline \hline
$u_v$ & 2 &~0.3056 $\pm$ 0.0023~&~0.2986$\pm$ 0.0029~&~0.285~&~0.304~&~0.2947~\\
      & 3 &~0.0871 $\pm$ 0.0009~&~0.0871$\pm$ 0.0011~&~0.082~&~0.087~&~0.0843~\\
      & 4 &~0.0330 $\pm$ 0.0004~&~0.0333$\pm$ 0.0005~&~0.032~&~0.033~&~0.0319~\\
\hline
$d_v$ & 2 &~0.1235 $\pm$ 0.0023~&~0.1239$\pm$ 0.0026~&~0.115~&~0.120~&~0.1129\\
      & 3 &~0.0298 $\pm$ 0.0008~&~0.0315$\pm$ 0.0008~&~0.028~&~0.028~&~0.0275\\
      & 4 &~0.0098 $\pm$ 0.0004~&~0.0105$\pm$ 0.0004~&~0.009~&~0.010~&~0.0092\\
\hline \hline
\end{tabular}
\normalsize
\vspace{2mm} \caption{\label{tab:table6}{\sf Comparison of low
order moments from our nonsinglet NNLO QCD analysis at $Q_0^2 =
4~\mbox{GeV}^2$ with the NNLO analysis BBG~\cite{Blumlein:2006be},
MRST04~\cite{MRST04}, A02~\cite{A02} and
A06~\cite{Alekhin:2006zm}. }}
\end{table*}

\begin{figure}
\begin{center}
\includegraphics[angle=0, width=9.0cm]{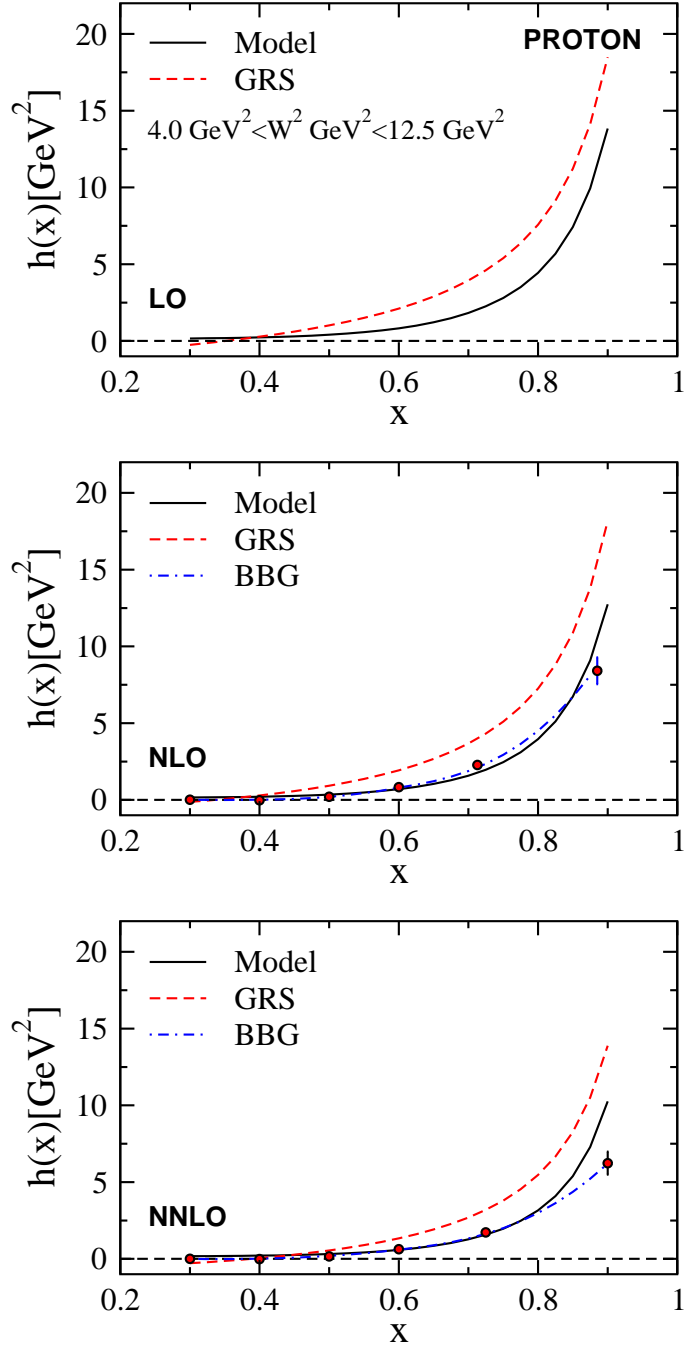}
\end{center}
{\sf \caption{\label{fig:7} The higher twist coefficient $h(x)$
for the proton data as a function of $x$  up to NNLO (solid line)
compared with results obtained by GRS (dashed line)
\cite{Gluck:2006yz} and BBG (dashed--dotted line)
\cite{Blumlein:2008kz}. }}
\end{figure}

To perform higher twist QCD analysis of the nonsinglet world data
up to NNLO, we consider the $Q^2 \geq 4 \GeV^2, 4<W^2 < 12.5
\GeV^2$ cuts. The number of data points in the above range for
proton and deuteron is 279 and  278, respectively.
 The extracted distributions for $h(x)$ up to NNLO are depicted in
Fig.(\ref{fig:7}) for the nonsinglet case considering scattering
off the proton target. According to our results the coefficient
$h(x)$ grows towards large $x$. Also in this figure HT
contributions have the tendency to decrease form LO to NLO, NNLO.
This effect was observed for the first time in the case of fits of
$F_3$ DIS $\nu N$ data in \cite{Kataev:1997nc} and then studied in
more detail in \cite{Kataev:1999bp,Kataev:2001kk}.

\begin{figure}
\begin{center}
\includegraphics[angle=0, width=9.0cm]{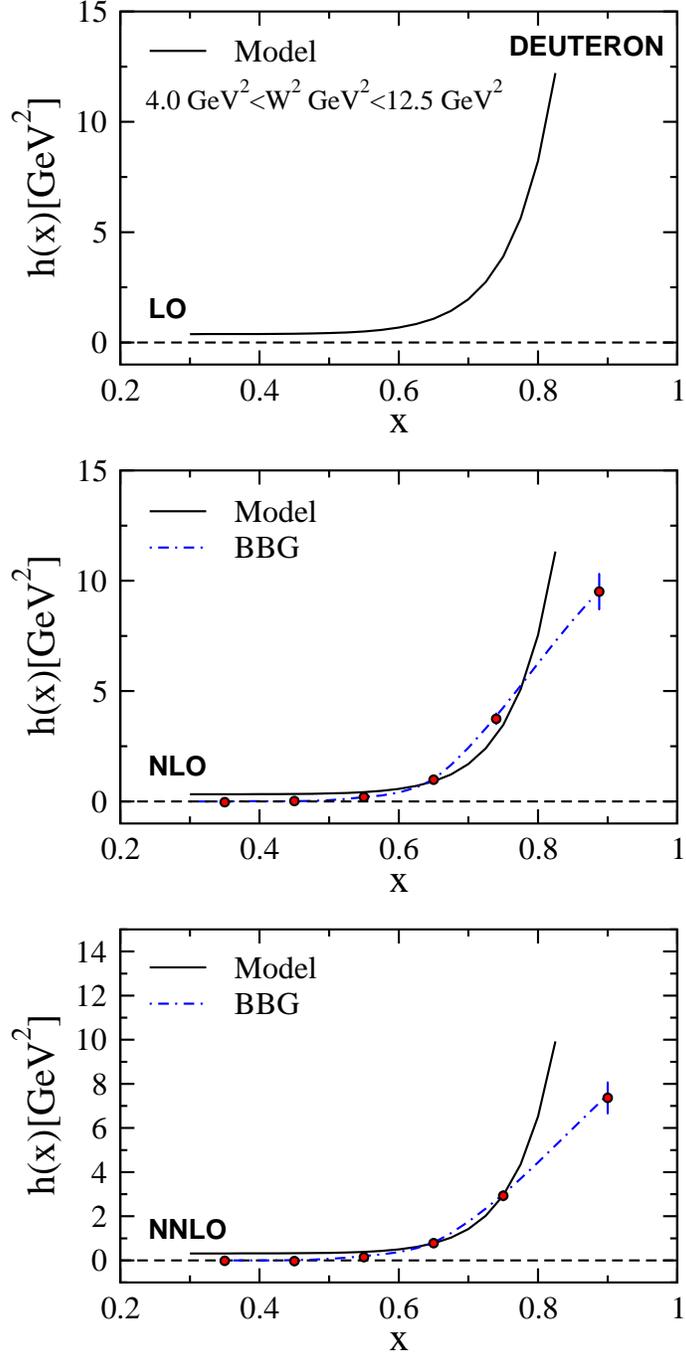}
\end{center}
{\sf \caption{\label{fig:8} The higher twist coefficient $h(x)$
for the deuteron data as a function of $x$ up to NNLO (solid line)
compared with results obtained by BBG (dashed--dotted line)
\cite{Blumlein:2008kz}.}}
\end{figure}

This similar effect was also observed in the fits of $F_2$ charge
lepton-nucleon DIS data
\cite{Yang:1999xg,Blumlein:2006be,Gluck:2006yz,Blumlein:2008kz}.
To compare, we also present the reported results of the early NNLO
analysis \cite{Blumlein:2008kz,Gluck:2006yz} in Fig.(\ref{fig:7}).
Note that the results for $h(x)$ in LO are not presented in the
BBG model \cite{Blumlein:2006be,Blumlein:2008kz}. In Ref.
\cite{Gluck:2006yz}, the functional form for $h(x)$ is chosen by
\begin{equation}
h(x) = a\left( \frac{x^b}{1-x} -c\right)\,.
\end{equation}
and it is possible to compare $h(x)$ results even in LO.
 Fig.(\ref{fig:8}) shows our results for $h(x)$ and for the
deuteron target up to NNLO. Also we compare the results for the
BBG model \cite{Blumlein:2008kz}. The same as the proton, HT
contributions for the deuteron have the tendency to decrease form
LO to NLO, NNLO. As seen from Fig.(\ref{fig:7}) and
Fig.(\ref{fig:8}) $h(x)$ is widely independent of the target
comparing the results for deeply inelastic scattering off protons
and deuterons. Our results in low-$x$ are also in good agreement
with \cite{Blumlein:2006be,Blumlein:2008kz}.

\section{Discussion}
We have performed a QCD analysis of the flavor nonsinglet
unpolarized deep--inelastic charged lepton--nucleon scattering
data to next--to--leading order and derived parameterizations of
valence quark distributions at a starting scale $Q_0^2$ together
with the QCD--scale $\Lambda_{\rm QCD}$ by using the Jacobi
polynomial expansions.

The analysis was performed using the Jacobi polynomials--method to
determine the parameters of the problem in a fit to the data. A
new aspect in comparison with previous analysis is that we
determine the parton densities and the QCD scale up to NNLO by
using  the Jacobi polynomial expansion method. The benefit of this
approach is the possibility to determine nonsinglet parton
distributions analytically and not numerically. In Ref.
\cite{Program:summary} we arrange the MATHEMATICA program to
extract $xu_v(x,Q^2)$ and $xd_v(x,Q^2)$.

In this paper the flavor asymmetric combination of light  parton
distributions $x(\overline{d}-\overline{u})$ of Eq.~(15) are fixed
at $Q_0^2=$ 4 GeV$^2$, as GRS \cite{Gluck:2006yz} and BBG
\cite{Blumlein:2004pr,Blumlein:2006be} applied, and gives a good
description of the Drell-Yan dimuon production data
\cite{Towell:2001nh}.
 The first clear evidence for the flavor
asymmetry of the nucleon sea in nature came from the analysis of
NMC at CERN \cite{Amaudruz:1991at}. In order to have the link with
NMC data, we want to study the compatibility of the
$x(\overline{d}-\overline{u})$ with the NMC result for the
Gottfried sum rule (GSR)\cite{Gottfried:1967kk}. This sum rule is
still actively discussed in problems of deep-inelastic scattering.
The GSR, $I_{GSR}$, can be expressed in terms of the parton
distribution functions as
\begin{eqnarray}
I_{GSR}(Q^2)&\equiv&\int_0^1\bigg[F_2^{lp}(x,Q^2)-F_2^{ln}(x,Q^2)\bigg]\frac{dx}{x}\nonumber
\\
&=&\int_0^1\bigg[\frac{1}{3}\bigg(u_v(x,Q^2)-d_v(x,Q^2)\bigg)+\frac{2}{3}
\bigg(\overline{u}(x,Q^2)-\overline{d}(x,Q^2)\bigg)\bigg]dx \nonumber\\
 &=& \frac{1}{3}+\frac{2}{3}\int_0^1
\bigg(\overline{u}(x,Q^2)-\overline{d}(x,Q^2)\bigg)dx~.
\label{GRS}
\end{eqnarray}
In the derivation of the above equation, the asymmetry of nucleon
sea was assumed. The NMC measurement \cite{Amaudruz:1991at}
implies at $Q^2=4$ GeV$^2$
\begin{eqnarray}
\int_0^1
(\overline{d}(x,Q^2)-\overline{u}(x,Q^2))dx=0.148\pm0.039~,
\end{eqnarray}
which was the first indication that there are more down antiquarks
in the proton than up antiquarks. On the other hand this value is
reported  $0.118\pm0.012$ at $Q^2=54$ GeV$^2$ \cite{E866}. Now it
is interesting to obtain this value  for the parametrization of
Eq.~(\ref{Asy}) which we used in our QCD analysis. By integration
of this distribution we obtain $\simeq0.1$ which is smaller than
the reported results in the literature.  However, the NMC
Collaboration gives the $I_{GSR}$ experimental value at $Q^2=4$
GeV$^2$ \cite{Amaudruz:1991at}
\begin{equation}
I_{GSR}^{exp}(Q^2=4~{\rm GeV}^2)=0.235\pm0.026~.
\end{equation}
By using Eq.~(\ref{GRS}) we obtain the GSR value  about 0.267 with
which the existing measurements are almost compatible within
error. It seems that although the value of $\int_0^1
(\overline{d}-\overline{u})dx$ is smaller than the values in the
literature, the parametrization of Eq.~(\ref{Asy}) can give a good
description of the E866 experimental data \cite{E866}. Also we
should notice that the GSR does not belong to the strict sum rules
in QCD and it is necessary to receive not only QCD corrections but
anomalous dimensions as well
\cite{Kataev:2007jz,Kataev:2003en,Kataev:2003xp,Broadhurst:2004jx,Broadhurst:2004kh,Kataev:2004wv}.

Now it is interesting to compare the NNLO theoretical QCD
theoretical prediction for the Gottfried sum rule
\cite{Broadhurst:2004jx} with NMC data. The recent step in this
direction was done in \cite{Abbate:2005ct}. According to this
paper we add the QCD two-loop correction to the Gottfried sum rule
and we refine the GSR value to about $0.12\%$. Also we obtain the
value of $I_{GSR}(0.004<x<0.8,4~\hbox{GeV}^2)=0.267$ which is well
compatible with the neural parametrization results, e.g.
$0.2281\pm 0.0437$ \cite{Abbate:2005ct} within errors.

In the QCD analysis we parameterized the strong coupling constant
$\alpha_s$ in terms of four massless flavors determining
$\Lambda_{\rm QCD}$. The LO, NLO, and NNLO results fitting the
data, are
\begin{eqnarray}
\Lambda_{\rm QCD}^{(4)\rm \overline{MS}} &=& 213.2 \pm 28
\; \mbox{MeV},~~{\tt LO}, \nonumber \\
\Lambda_{\rm QCD}^{(4)\rm \overline{MS}} &=& 263.8 \pm 30 \;
\mbox{MeV},~~{\tt NLO}, \nonumber \\
\Lambda_{\rm QCD}^{(4)\rm \overline{MS}} &=& 239.9 \pm 27 \;
\mbox{MeV},~~{\tt NNLO},
\end{eqnarray}
These results can be expressed in terms of $\alpha_s(M_Z^2)$:
\begin{eqnarray}
\alpha_s(M_Z^2) = 0.1281 \pm 0.0028,~~{\tt LO},  \nonumber \\
\alpha_s(M_Z^2) = 0.1149 \pm 0.0021,~~{\tt NLO}, \nonumber \\
\alpha_s(M_Z^2) = 0.1131 \pm 0.0019,~~{\tt NNLO}.
\end{eqnarray}
Note that in above results we use the matching between $n_f$ and
$n_{f+1}$ flavor couplings calculated in Ref.
\cite{Chetyrkin:1997sg}. To be capable to compare with other
measurement of $\Lambda_{\rm QCD}$ we adopt this prescription.

 The $\alpha_s(M_Z^2)$ values can be
compared with results from other QCD analysis of inclusive
deep--inelastic scattering data in NLO
%
%
\renewcommand{\arraystretch}{0.9}
\normalsize
\begin{center}
\begin{tabular}{rcllll}
& & & & & \\

 A02 \cite{A02}:   & & $\alpha_s(M_Z^2)$=0.1171 & $\pm$0.0015 &               \\
 ZEUS \cite{ZEUS_Ch}:  & & $\alpha_s(M_Z^2)$=0.1166 & $\pm$0.0049 &    \\
 H1 \cite{H1}:    & & $\alpha_s(M_Z^2)$=0.1150 & $\pm$0.0017 &

 \\
 BCDMS \cite{BCDMS}: & & $\alpha_s(M_Z^2)$=0.110  & $\pm$0.006 &                          \\
GRS \cite{Gluck:2006yz}: & &  $\alpha_s(M_Z^2)$=0.112  &  &
\\
CTEQ6 \cite{CTEQ_P}:&~~ & $\alpha_s(M_Z^2)$=0.1165 & $\pm$0.0065 &                           \\
 MRST03 \cite{MRST03}:&~~ & $\alpha_s(M_Z^2)$=0.1165 & $\pm$0.0020 &
\\
BBG \cite{Blumlein:2006be}: & &  $\alpha_s(M_Z^2)$=0.1148  & $\pm$
0.0019 &
\\
KK05 \cite{Krivokhizhin:2005pt}: & &  $\alpha_s(M_Z^2)$=0.1153  &
$\pm$ 0.0013($stat$)\\
&&&$\pm$0.0022($syst$) $\pm$0.0012($norm$) &
\\
 BB (pol)\cite{BB02}:   &  & $\alpha_s(M_Z^2)$=0.113 & $\pm$0.004 &
\\
AK (pol) \cite{Atashbar Tehrani:2007be}: & &
$\alpha_s(M_Z^2)$=0.1141  & $\pm$ 0.0036 &
\end{tabular}
\end{center}
\normalsize

The NNLO values of $\alpha_s(M_Z^2)$ can also be compared with
results from other QCD analysis

\normalsize
\begin{center}
\begin{tabular}{rcllll}
 & & & & & \\
A02   \cite{A02}:      & & $\alpha_s(M_Z^2)$=0.1143 & $\pm$0.0014 &   \\
GRS  \cite{Gluck:2006yz}:&  & $\alpha_s(M_Z^2)$=0.111 &  &
\\
 MRST03 \cite{MRST03}:     & &$\alpha_s(M_Z^2)$= 0.1153 & $\pm$0.0020 &    \\
 SY01(ep)  \cite{SY}:  & & $\alpha_s(M_Z^2)$=0.1166 & $\pm$0.0013 &                \\
 SY01($\nu$N) \cite{SY}:& & $\alpha_s(M_Z^2)$=0.1153 & $\pm$0.0063 &                \\
A06 \cite{Alekhin:2006zm}: & &  $\alpha_s(M_Z^2)$=0.1128 & $\pm$
0.0015 &
\\
BBG \cite{Blumlein:2006be}:  & & $\alpha_s(M_Z^2)$=0.1134  & \footnotesize{$\begin{array}{c} +0.0019 \\
-0.0021 \end{array}$} &\\
BM07   \cite{Brooks:2006wh}:      & & $\alpha_s(M_Z^2)$=0.1189 &
$\pm$0.0019 &
\\
KPS00($\nu N$) \cite{Kataev:1999bp}:      & &
$\alpha_s(M_Z^2)$=0.118 &$\pm$ 0.002 ~($stat$)$\pm$ 0.005~($syst$) \\
&&& $\pm$
0.003~($theory$) &  \\
KPS03($\nu N$) \cite{Kataev:2001kk}:      & &
$\alpha_s(M_Z^2)$=0.119 &$\pm$ 0.002 ~($stat$)$\pm$ 0.005~($syst$)\\
&&&  $\pm$ 0.002 ~($threshold$)
$^{+0.004}_{-0.002}$~ ($scale$) &  \\
\end{tabular}
\end{center}
\normalsize

\noindent and with the value of the current world average
\begin{eqnarray}
 \alpha_s(M_Z^2) = 0.1189 \pm 0.0010~,
\end{eqnarray}
which has been extracted in \cite{Bethke:2006ac} recently.

We hope our results of QCD analysis of structure functions in
terms of Jacobi polynomials
 could be able to describe more complicated hadron structure functions. We also
hope to be able to consider the N$^3$LO corrections and massive
quark contributions by using the structure function expansion in
terms of the Jacobi polynomials.

\begin{acknowledgments}
We are especially grateful to G. Altarelli and J. Bl\"umlein   for
guidance and critical remarks. A.N.K. is grateful to S. Albino for
useful discussions. We would like to thank M. Ghominejad and Z.
Karamloo for reading the manuscript of this paper. A.N.K. thanks
Semnan University for partial financial support of this project.
We acknowledge the Institute for Studies in Theoretical Physics
and Mathematics (IPM) for financially supporting this project.
\end{acknowledgments}

\newpage
\newpage
\newpage
\newpage

\newpage
\newpage

\normalsize
\normalsize

\clearpage

\bibliography{apssamp}

\begin{thebibliography}{99}
%


\bibitem{Altarelli:2008fd}
  G.~Altarelli,
  arXiv:0804.4147 [hep-ph].

\bibitem{Altarelli:2008aj}
  G.~Altarelli, R.~D.~Ball and S.~Forte,
  Nucl.\ Phys.\  B {\bf 799}, 199 (2008)
  [arXiv:0802.0032 [hep-ph]].


\bibitem{Altarelli:2008xp}
  G.~Altarelli, R.~D.~Ball and S.~Forte,
  PoS {\bf RADCOR2007}, 028 (2007)
  [arXiv:0802.0968 [hep-ph]].


\bibitem{Atashbar Tehrani:2007be}
  S.~Atashbar Tehrani and A.~N.~Khorramian,
  JHEP {\bf 0707}, 048 (2007)
  [arXiv:0705.2647 [hep-ph]].

\bibitem{Khorramian:2006wg}
  A.~N.~Khorramian and S.~Atashbar Tehrani,
  JHEP {\bf 0703}, 051 (2007)
  [arXiv:hep-ph/0610136].


\bibitem{parisi}
G. Parisi and N. Sourlas,
\newblock{\it Nucl. Phys.} {\bf B151} (1979) 421;
\\ I. S. Barker, C. B. Langensiepen and  G. Shaw,
\newblock{\it Nucl. Phys.} {\bf B186} (1981) 61.

\bibitem{Barker}
I. S. Barker, B. R. Martin and G. Shaw,
\newblock{\it Z. Phys.} {\bf C19} (1983)  147;
\\I. S. Barker and B. R. Martin,
\newblock{\it Z. Phys.} {\bf C24} (1984) 255;
\\S. P. Kurlovich, A. V. Sidorov and N. B. Skachkov,
\newblock JINR Report E2-89-655, Dubna, 1989.


\bibitem{Krivokhizhin:1987rz}
  V.~G.~Krivokhizhin, S.~P.~Kurlovich, V.~V.~Sanadze, I.~A.~Savin, A.~V.~Sidorov and N.~B.~Skachkov,
  Z.\ Phys.\ C {\bf 36} (1987) 51.


\bibitem{Krivokhizhin:1990ct}
  V.~G.~Krivokhizhin {\it et al.},
  Z.\ Phys.\ C {\bf 48}, 347 (1990).


\bibitem{Chyla:1986eb}
  J.~Chyla and J.~Rames,
  Z.\ Phys.\  C {\bf 31} (1986) 151.




\bibitem{Barker:1980wu}
  I.~S.~Barker, C.~S.~Langensiepen and G.~Shaw,
  Nucl.\ Phys.\  B {\bf 186} (1981) 61.



\bibitem{Kataev:1997nc}
  A.~L.~Kataev, A.~V.~Kotikov, G.~Parente and A.~V.~Sidorov,
  Phys.\ Lett.\ B {\bf 417}, (1998) 374
  [arXiv:hep-ph/9706534].


\bibitem{Kataev:1998ce}
  A.~L.~Kataev, G.~Parente and A.~V.~Sidorov,
  arXiv:hep-ph/9809500.






\bibitem{Kataev:1999bp}
  A.~L.~Kataev, G.~Parente and A.~V.~Sidorov,
  Nucl.\ Phys.\ B {\bf 573}, (2000) 405
  [arXiv:hep-ph/9905310].




\bibitem{Kataev:2001kk}
  A.~L.~Kataev, G.~Parente and A.~V.~Sidorov,
  Phys.\ Part.\ Nucl.\  {\bf 34},  (2003) 20
   [arXiv:hep-ph/0106221]; \\A.~L.~Kataev, G.~Parente and A.~V.~Sidorov,
Nucl.\ Phys.\ Proc.\ Suppl.\  {\bf 116} (2003) 105
[arXiv:hep-ph/0211151].

\bibitem{Khorramian:2007zz}
  A.~N.~Khorramian, S.~A.~Tehrani and M.~Ghominejad,
  Acta Phys.\ Polon.\  B {\bf 38}, 3551 (2007).

\bibitem{Leader:1997kw}
  E.~Leader, A.~V.~Sidorov and D.~B.~Stamenov,
  Int.\ J.\ Mod.\ Phys.\  A {\bf 13}, 5573 (1998)
  [arXiv:hep-ph/9708335].


\bibitem{Khorramian:2007gu}
  A.~N.~Khorramian and S.~A.~Tehrani,
  arXiv:0712.2373 [hep-ph].


\bibitem{Khorramian:2007zza}
  A.~N.~Khorramian and S.~Atashbar Tehrani,
  AIP Conf.\ Proc.\  {\bf 915}, 420 (2007).


\bibitem{Mirjalili:2007ep}
  A.~Mirjalili, A.~N.~Khorramian and S.~Atashbar-Tehrani,
  Nucl.\ Phys.\ Proc.\ Suppl.\  {\bf 164}, 38 (2007).

\bibitem{Mirjalili:2006hf}
  A.~Mirjalili, S.~Atashbar Tehrani and A.~N.~Khorramian,
  Int.\ J.\ Mod.\ Phys.\  A {\bf 21}, 4599 (2006)
  [arXiv:hep-ph/0608224].


\bibitem{ref3} W.L.~van Neerven, A.~Vogt, {\it Nucl.~Phys.}
               {\bf B568}, 263 (2000)
\bibitem{ref4} W.L.~van Neerven, A.~Vogt, {\it Nucl.~Phys.}
               {\bf B588}, 345 (2000) and arXiv:hep-ph/0006154
               (corrected)
\bibitem{ref5} J.~Bl\"umlein, A.~Vogt,
               {\it Phys.~Rev.} {\bf D58}, 014020 (1998).
\bibitem{Gluck:2006pm}
  M.~Gluck, C.~Pisano and E.~Reya,
  Eur.\ Phys.\ J.\  C {\bf 50}, 29 (2007)
  [arXiv:hep-ph/0610060].

\bibitem{Vermaseren:2005qc}
J.~A.~M.~Vermaseren, A.~Vogt and S.~Moch,
Nucl.\ Phys.\ B {\bf 724} (2005) 3 [arXiv:hep-ph/0504242].


\bibitem{E866} R.S.~Towell et al., E866 Collab.,
               {\it Phys.~Rev.} {\bf D64} (2001) 052002.

\bibitem{ref19} A.D.~Martin  et al., {\it Eur.~Phys.~J.} {\bf C23}
               (2002) 73.
\bibitem{Blumlein:2004pr} J.~Bl\"umlein, H,~B\"ottcher, and A.~Guffanti,
               {\it Nucl.~Phys.~B} (Proc.~Suppl.)
               {\bf 135} (2004) 152.

\bibitem{Blumlein:2006be}
  J.~Blumlein, H.~Bottcher and A.~Guffanti,
  Nucl.\ Phys.\  B {\bf 774}, 182 (2007)
  [arXiv:hep-ph/0607200].

\bibitem{FP}
W.~Furmanski and R.~Petronzio,
Z.\ Phys.\ C {\bf 11} (1982) 293.


\bibitem{NS2}
W.~L.~van Neerven and E.~B.~Zijlstra,
Phys.\ Lett.\ B {\bf 272} (1991) 127;\\
E.~B.~Zijlstra and W.~L.~van Neerven,
Nucl.\ Phys.\ B {\bf 383} (1992) 525.

\bibitem{Tarasov:1980au}
  O.~V.~Tarasov, A.~A.~Vladimirov and A.~Y.~Zharkov,
  Phys.\ Lett.\  B {\bf 93}, 429 (1980).


\bibitem{Larin:1993tp}
  S.~A.~Larin and J.~A.~M.~Vermaseren,
  Phys.\ Lett.\  B {\bf 303}, 334 (1993)
  [arXiv:hep-ph/9302208].

\bibitem{Vogt:2004ns}
  A.~Vogt,
  Comput.\ Phys.\ Commun.\  {\bf 170}, 65 (2005)
  [arXiv:hep-ph/0408244].


\bibitem{BCDMS}
A.C.~Benvenuti {\it et al.} [BCDMS Collaboration],
Phys.\ Lett.\ B {\bf 237} (1990) 592;
\\
A.C.~Benvenuti {\it et al.} [BCDMS Collaboration], Phys. Lett.
{\bf B223} (1989) 485; Phys. Lett. {\bf B237} (1990) 592.
\\
A.C.~Benvenuti {\it et al.} [BCDMS Collaboration],
Phys.\ Lett.\ B {\bf 237} (1990) 599.
%
\bibitem{SLAC}
L.~W.~Whitlow, E.~M.~Riordan, S.~Dasu, S.~Rock and A.~Bodek, %
Phys.\ Lett.\ B {\bf 282} (1992). %
%
\bibitem{NMC}
M.~Arneodo {\it et al.}  [New Muon Collaboration],
Nucl.\ Phys.\ B {\bf 483} (1997) 3 [arXiv:hep-ph/9610231].
%
\bibitem{H1}
C.~Adloff {\it et al.}  [H1 Collaboration],
Eur.\ Phys.\ J.\ C {\bf 21} (2001) 33
[arXiv:hep-ex/0012053];\\
C.~Adloff {\it et al.}  [H1 Collaboration],
Eur.\ Phys.\ J.\ C {\bf 30} (2003) 1 [arXiv:hep-ex/0304003].
%
\bibitem{ZEUS}
J.~Breitweg {\it et al.}  [ZEUS Collaboration],
Eur.\ Phys.\ J.\ C {\bf 7} (1999) 609
[arXiv:hep-ex/9809005];\\
S.~Chekanov {\it et al.}  [ZEUS Collaboration],
Eur.\ Phys.\ J.\ C {\bf 21} (2001) 443 [arXiv:hep-ex/0105090].

\bibitem{Stump:2001gu}
  D.~Stump {\it et al.},
  Phys.\ Rev.\  D {\bf 65}, 014012 (2002)
  [arXiv:hep-ph/0101051].


\bibitem{MINUIT}
F.~James, CERN Program Library, Long Writeup D506 (MINUIT).


\bibitem{Georgi:1976ve}
H.~Georgi and H.~D.~Politzer,
Phys.\ Rev.\ D {\bf 14} (1976) 1829.


\bibitem{Gluck:2006yz}
  M.~Gluck, E.~Reya and C.~Schuck,
  Nucl.\ Phys.\  B {\bf 754}, 178 (2006)
  [arXiv:hep-ph/0604116].


\bibitem{Alekhin:2005gq}
  S.~Alekhin,
  JETP Lett.\  {\bf 82}, 628 (2005)
  [Pisma Zh.\ Eksp.\ Teor.\ Fiz.\  {\bf 82}, 710 (2005)]
  [arXiv:hep-ph/0508248].


\bibitem{Martin:2007bv}
  A.~D.~Martin, W.~J.~Stirling, R.~S.~Thorne and G.~Watt,
  Phys.\ Lett.\  B {\bf 652}, 292 (2007)
  [arXiv:0706.0459 [hep-ph]].

  \bibitem{MRST04}
A.~D.~Martin, R.~G.~Roberts, W.~J.~Stirling and R.~S.~Thorne,
Phys.\ Lett.\ B {\bf 604} (2004) 61 [arXiv:hep-ph/0410230].

\bibitem{A02}
S.~Alekhin,
Phys.\ Rev.\ D {\bf 68} (2003) 014002 [arXiv:hep-ph/0211096].

\bibitem{Alekhin:2006zm}
  S.~Alekhin, K.~Melnikov and F.~Petriello,
  Phys.\ Rev.\  D {\bf 74}, 054033 (2006)
  [arXiv:hep-ph/0606237].

\bibitem{Yang:1999xg}
  U.~K.~Yang and A.~Bodek,
  Eur.\ Phys.\ J.\  C {\bf 13}, 241 (2000)
  [arXiv:hep-ex/9908058].

\bibitem{Blumlein:2008kz}
  J.~Blumlein and H.~Bottcher,
  Phys.\ Lett.\  B {\bf 662}, 336 (2008)
  [arXiv:0802.0408 [hep-ph]].


\bibitem{Program:summary}
Program summary URL: http://particles.ipm.ir/QCD.htm.

\bibitem{Towell:2001nh}
  R.~S.~Towell {\it et al.}  [FNAL E866/NuSea Collaboration],
  Phys.\ Rev.\  D {\bf 64}, 052002 (2001)
  [arXiv:hep-ex/0103030].


\bibitem{Amaudruz:1991at}
  P.~Amaudruz {\it et al.}  [New Muon Collaboration],
  Phys.\ Rev.\ Lett.\  {\bf 66}, 2712 (1991);
M.~Arneodo {\it et al.}  [New Muon Collaboration],
  Phys.\ Rev.\  D {\bf 50}, 1 (1994);
  M.~Arneodo {\it et al.}  [New Muon Collaboration],
  Nucl.\ Phys.\  B {\bf 487}, 3 (1997)
  [arXiv:hep-ex/9611022].

\bibitem{Gottfried:1967kk}
K.~Gottfried,
Phys.\ Rev.\ Lett.\  {\bf 18} (1967) 1174.


\bibitem{Kataev:2003en}
  A.~L.~Kataev and G.~Parente,
  Phys.\ Lett.\  B {\bf 566}, 120 (2003)
  [arXiv:hep-ph/0304072].


\bibitem{Kataev:2003xp}
  A.~L.~Kataev,
  arXiv:hep-ph/0311091.

\bibitem{Broadhurst:2004jx}
  D.~J.~Broadhurst, A.~L.~Kataev and C.~J.~Maxwell,
  Phys.\ Lett.\  B {\bf 590}, 76 (2004)
  [arXiv:hep-ph/0403037].

\bibitem{Broadhurst:2004kh}
  D.~J.~Broadhurst, A.~L.~Kataev and C.~J.~Maxwell,
  arXiv:hep-ph/0410058.

\bibitem{Kataev:2004wv}
  A.~L.~Kataev,
  Phys.\ Part.\ Nucl.\  {\bf 36}, S168 (2005)
  [arXiv:hep-ph/0412369].


\bibitem{Kataev:2007jz}
  A.~L.~Kataev,
  PoS A {\bf CAT2007}, 072 (2007)
  [arXiv:0707.2855 [hep-ph]].
















\bibitem{Abbate:2005ct}
  R.~Abbate and S.~Forte,
  Phys.\ Rev.\  D {\bf 72}, 117503 (2005)
  [arXiv:hep-ph/0511231].

\bibitem{Chetyrkin:1997sg}
K.~G.~Chetyrkin, B.~A.~Kniehl and M.~Steinhauser,
, Phys.\ Rev.\ Lett.\  {\bf 79} (1997) 2184
[arXiv:hep-ph/9706430]; S.~Bethke,
, J.\ Phys.\ G {\bf 26} (2000) R27 [arXiv:hep-ex/0004021];
W.~Bernreuther and W.~Wetzel,
  Nucl.\ Phys.\  B {\bf 197}, 228 (1982)
  [Erratum-ibid.\  B {\bf 513}, 758 (1998)].






\bibitem{ZEUS_Ch}
S.~Chekanov {\it et al.}  [ZEUS Collaboration],
Phys.\ Rev.\ D {\bf 67} (2003) 012007 [arXiv:hep-ex/0208023].




\bibitem{CTEQ_P}
J.~Pumplin, D.~R.~Stump, J.~Huston, H.~L.~Lai, P.~Nadolsky and
W.~K.~Tung,
JHEP {\bf 0207} (2002) 012 [arXiv:hep-ph/0201195].

\bibitem{MRST03}
A.~D.~Martin, R.~G.~Roberts, W.~J.~Stirling and R.~S.~Thorne,
arXiv:hep-ph/0307262.

\bibitem{Krivokhizhin:2005pt}
  V.~G.~Krivokhizhin and A.~V.~Kotikov,
  Phys.\ Atom.\ Nucl.\  {\bf 68}, 1873 (2005)
  [Yad.\ Fiz.\  {\bf 68}, 1935 (2005)];



\bibitem{BB02}
J.~Bl\"umlein and H.~B\"ottcher,
Nucl.\ Phys.\ B {\bf 636} (2002) 225 [arXiv:hep-ph/0203155].





\bibitem{SY}
J.~Santiago and F.~J.~Yndurain,
Nucl.\ Phys.\ B {\bf 611} (2001) 447 [arXiv:hep-ph/0102247].


\bibitem{Brooks:2006wh}
  P.~M.~Brooks and C.~J.~Maxwell,
  Nucl.\ Phys.\  B {\bf 780}, 76 (2007)
  [arXiv:hep-ph/0610137].


\bibitem{Bethke:2006ac}
  S.~Bethke,
  Prog.\ Part.\ Nucl.\ Phys.\  {\bf 58}, 351 (2007)
  [arXiv:hep-ex/0606035].



\end{thebibliography}

\end{document}